\documentclass[11pt]{article}

\usepackage[margin=1in]{geometry}
\usepackage[utf8]{inputenc}

\usepackage{mathtools}
\usepackage{amssymb}
\usepackage{booktabs}
\usepackage{multirow}
\usepackage{array}
\usepackage{algorithm}
\usepackage{algpseudocode}
\usepackage{enumitem}
\usepackage{subcaption}
\usepackage{float}
\usepackage{lscape}
\usepackage{rotating}

\usepackage[numbers]{natbib}

\usepackage{graphicx}
\usepackage[table]{xcolor}

\graphicspath{{./}}

\definecolor{ForestGreen}{rgb}{0.13,0.55,0.13}
\definecolor{BrickRed}{rgb}{0.80,0.25,0.33}

\definecolor{tableheadgray}{rgb}{0.937,0.937,0.937}
\definecolor{nashblue}{rgb}{0.843,0.965,0.988}
\definecolor{headerblue}{rgb}{0.192,0.400,1.000}
\definecolor{rowlabelred}{rgb}{0.796,0.000,0.000}

\def\PST{\textit{PST}}
\def\Away{\textit{Away}}

\usepackage{hyperref}
\hypersetup{
    colorlinks=true,
    linkcolor=blue,
    citecolor=blue,
    urlcolor=blue
}

\title{Hybrid Human--Agent Social Dilemmas in Energy Markets}

\author{Isuri Perera, Frits de Nijs, and Julian Garcia\\
Department of Data Science and AI, Faculty of Information Technology\\
Monash University, Melbourne, Australia\\
}

\date{}

\begin{document}

\maketitle

\begin{abstract}
In hybrid populations where humans delegate strategic decision-making to autonomous agents, understanding when and how cooperative behaviors can emerge remains a key challenge.
We study this problem in the context of energy load management: consumer agents schedule their appliance use under demand-dependent pricing.
This structure can create a social dilemma where everybody would benefit from coordination, but in equilibrium agents often choose to incur the congestion costs that cooperative turn-taking would avoid.
To address the problem of coordination, we introduce artificial agents that use globally observable signals to increase coordination. Using evolutionary dynamics, and reinforcement learning experiments, we show that artificial agents can shift the learning dynamics to favour coordination outcomes.
An often neglected problem is partial adoption: what happens when the technology of artificial agents is in the early adoption stages?
We analyze mixed populations of adopters and non-adopters, demonstrating that unilateral entry is feasible: adopters are not structurally penalized, and partial adoption can still improve aggregate outcomes. However, in some parameter regimes, non-adopters may benefit disproportionately from the cooperation induced by adopters. This asymmetry, while not precluding beneficial entry, warrants consideration in deployment, and highlights strategic issues around the adoption of AI technology in multiagent settings.
\end{abstract}


\section{Introduction}

As autonomous agents increasingly act on behalf of human principals in strategic settings, understanding the evolution of sociality in hybrid human--agent populations becomes essential~\cite{DafoeEtAl2021CooperativeAI,BarfussEtAl2025Collective}.
In domains ranging from energy markets to traffic routing, humans can delegate scheduling and bidding decisions to algorithmic agents.
These agents interact repeatedly, and the strategies or policies they employ, whether cooperative or competitive, shape collective outcomes.
In this context, the decision to adopt delegation, and the design of the delegated agent, can itself have strategic consequences.
When some individuals in a population adopt artificial agents while others do not, the resulting mixed population exhibits dynamics that can be analyzed using evolutionary game theory~\cite{GuoEtAl2023FacilitatingCooperationHybrid,GuoShen2025PowerAsymmetryBots}.
We argue that this perspective connects the practical question of technology adoption to the formal study of how behavioral strategies spread or decline in populations where humans and artificial agents interact.

An important use case is electricity markets.
Common electricity generators vastly vary in their capacity to deliver power and cost of operation.
As a result, the cost of electricity is not constant but depends on the number and types of generators that are called on to produce at a given moment in time~\cite{strbac2008demand}.
The advent of renewable energy has introduced a valuable alternative with near-zero operational costs, but the generation is limited by the availability of the energy source.
Because of this, electricity prices now fluctuate significantly with the time-of-day and the aggregate demand of consumers.

To optimize electricity usage, rational consumers should schedule their electric appliances during periods of lower costs.
However, if a significant number of consumers adopt this strategy, this will increase overall demand, triggering the activation of other expensive generators and, consequently, raising overall costs.
Addressing this challenge demands coordination among consumers, giving rise to the multi-agent problem of demand side load management (DSLM).

We approach this problem through the lens of evolutionary game theory.
While traditional game-theoretic analysis can identify equilibria, it is on its own silent on which equilibrium is picked or learned by autonomous agents~\cite{GarciaTraulsen2025Picking}.
Through a theoretically tractable repeated game, we aim to attain insights into the dynamics, strategic interactions, and the availability of socially beneficial and non-beneficial equilibria.
Evolutionary dynamics reveals the plausibility of equilibria learning agents will converge to.
This provides a bridge between the practical application and the formal study of how artificial agents may influence learning dynamics.

We consider the case where human agents can delegate strategic behavior to artificial agents, resulting in hybrid populations: some decisions are made by humans directly, others by autonomous agents on their behalf.
Given that consumers have heterogeneous preferences for operation times, it is not trivial to understand others' objectives, making coordination particularly difficult at scale.
In the literature, many approaches have been proposed, with some considering demand-independent pricing rules to create a stationary environment.
However, these approaches often overlook the complexities of multi-agent settings and run the risk of simply shifting peak demand rather than effectively curbing it~\cite{van2013randomized}.

The most successful settings for demand side load control typically involve a central agent, often the utility company, playing a pivotal role by providing guidance through price signals.
These price signals serve as an additional congestion cost to make cheap periods less attractive to agents.
While most such partially centralized approaches result in lesser aggregate prices they tend to ignore the impact on individual consumers, and rely primarily on the central agent.

We capture and formalize the dynamics of the DSLM problem and specifically investigate a fully decentralized approach to scheduling individual consumer appliances.

Much of the literature on artificial agent design for multi-agent coordination assumes universal adoption or mandated participation.
In practice, however, adoption is voluntary and gradual.
The question of whether early adopters of agent delegation will be disadvantaged, whether partial adoption can still yield benefits, and whether ``free-riding'' by non-adopters undermines the adoption of the technology, has received insufficient attention.
We call delegated agents \emph{entry resilient} if unilateral adoption does not make the adopter worse off when facing non-adopters; that is, given the chance to adjust the payoffs, it is advantageous for a cost-minimizing agent to adopt this approach, irrespective of whether the opponents choose to adopt or not.

In a free market setting, consumers will have to be convinced of the value of a load-managing device, with the option to choose not to participate in the scheme if they prefer.
Therefore, we assume agents that have the option to adopt a payoff shaping scheme that does not incur additional costs.

Given the extensive body of literature on demand response~\cite{vazquez2019reinforcement,PereraKamalaruban2021RLEnergyReview}, we focus specifically on multi-agent research related to DSLM where the price is demand-dependent.
The majority of studies in this category concentrate on distinct energy needs such as heating, ventilation, and air conditioning (HVAC), domestic hot water (DHW), appliances, and electric vehicles (EVs), focusing on purely competitive settings such as energy markets or only the competitive aspects of demand fulfillment~\cite{dauer2013market,dusparic2013multi,vaya2014optimal,sun2015learning,kim2015dynamic,bahrami2017online}.
In this domain, it is often assumed that a central controller is available to guide agents towards globally desirable outcomes via price signals~\cite{miri2023demand,he2018fast,CharbonnierMorstynMcCulloch2022ScalableMARL}.
Such systems require the participation of all consumers and often result in a biased system if private information is not shared with the central agent.
This also allows individuals to exploit the system by providing false information.
Moving to research that focuses on the collaborative aspect of DSLM, many cooperative approaches in literature either have a service provider price signaling agents or agents sharing information partially in their attempts to reduce costs~\cite{zhu2014distributed,zhang2017deep,kofinas2018fuzzy,jiang2018multiple}.
Most other collaborative research focuses on multi-objective learning where agents either transfer knowledge or model others~\cite{taylor2014accelerating,marinescu2015p,dusparic2015maximizing}.
These approaches are susceptible to privacy concerns and might not consistently align with the inherently self-interested nature of the agents.

Our approach stands apart from the existing literature due to our focus on a fully decentralized system where agents coordinate using only globally observable signals (aggregate demand and prices), without preference sharing, and with explicit analysis of entry feasibility and partial adoption in mixed populations.

The remainder of this paper is organized as follows. Section~\ref{sec:social-dilemma} presents the energy load management domain and demonstrates that it constitutes a social dilemma; this is done by using a centralized benchmark and a decentralized RL baseline. Section~\ref{sec:minimal-game} abstracts this to a minimal stage game and analyzes the repeated game to identify when cooperative turn-taking becomes feasible.
Section~\ref{sec:evolutionary} uses two-population replicator dynamics and Monte Carlo simulations to characterize when cooperation emerges versus when populations converge to non-cooperative equilibria.
Section~\ref{sec:delegation} introduces artificial agents with an intrinsic reward scheme based on globally observable signals that shifts evolutionary basins toward cooperation.
Section~\ref{sec:entry} analyzes entry feasibility and adoption resilience in mixed populations.
Section~\ref{sec:discussion} concludes.

\section{Energy load management as a social dilemma}
\label{sec:social-dilemma}

The Demand Side Load Control (DSLC) problem is usually formulated by  representing the utility provider, consumers, and appliances. Here, we follow the model presented by He et al.~\cite{he2018fast}. Each consumer has multiple appliances with parameters that capture the characteristics of each appliance as well as user preferences associated with their usage.
Key parameters include: appliance duration $d_{i,j}$, power requirement $w_{i,j}$, Preferred Start Time ($PST$), Earliest Start Time ($EST$), Latest Start Time ($LST$), and an inconvenience factor $\eta_{i,j}$ quantifying the cost of deviating from the preferred start time (see Appendix~\ref{app:details} for the full formulation).

The utility provider's pricing mechanism operates based on a pre-defined price table with step functions that increase with aggregate demand. Given an aggregate demand $P_t$ for planning slot $t$, the total price is calculated via a step function~$C_t(P_t)$ where higher demand bands incur higher per-unit costs. The total cost of consumer $i$ combines energy cost and inconvenience:
\begin{equation}
    C_{i} = \sum\limits_{t=0}^h (T_{i,t}) + I_{i}
\end{equation}
where $T_{i,t}$ is the energy cost for slot $t$ and $I_i$ is the total inconvenience from deviating from preferred start times.

To establish a benchmark, we model the centralized DSLC problem using MiniZinc~\cite{nethercote2007minizinc} and solve it to optimality. The aim is to calculate start times for each appliance to minimize the total system cost. For the decentralized solution, we implement a reinforcement learning algorithm where each consumer (agent) learns to select appliance start times based on observed price information~\cite{NweyeEtAl2022CityLearnChallenge}.
Full details of the RL implementation, including state representation, policy gradient updates, and hyperparameters, are provided in Appendix~\ref{app:rl}.

Figure~\ref{fig:CvD} illustrates the gap between centralized and decentralized outcomes.
Notably, substantial deviations from the optimal outcomes can be observed when agents learn with simple rewards.
The maximum decentralized cost reported for the population with 6 agents is more than double the maximum centrally optimal cost observed. This discrepancy indicates the potential for incurring significantly higher costs even when optimization is carried out to reduce individual costs.

\begin{figure}[ht]
    \centering
    \includegraphics[width=0.7\textwidth]{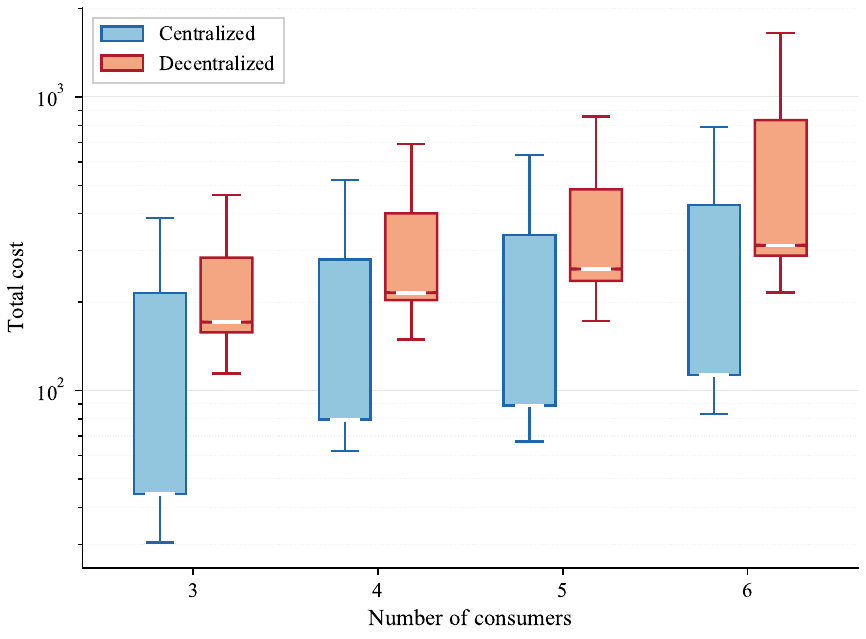}
    \caption{Centralized vs Decentralized outcomes for 130 testing days across different population sizes. Selfish agents learn subobtimal policies.}
    \label{fig:CvD}
\end{figure}

To illustrate the social dilemma structure, we examine a simplified scenario with two identical consumers who each aim to schedule the operation of a single appliance.
Both consumers share a Preferred Start Time of slot 2, with appliances requiring 5~kW of power over 10 hours. The price table comprises two pricing slots (first and second half of the day), each with a step function: \$5/kWh for the first 5~kW of demand, escalating to \$10/kWh thereafter. Both consumers have an inconvenience cost of $\eta = \$30$ per hour.

When both consumers schedule at their PST (slot 2), the simultaneous appliance activations lead to an aggregate demand of 10~kW, exceeding the 5~kW threshold. This results in the unit price escalating to \$10/kWh on average, yielding a total cost of \$500 + \$0 (energy plus discomfort) for each consumer and an aggregate system cost of \$1000. This situation is also the Nash Equilibrium of this simplified scenario, as unilaterally deviating by one step increases the costs to \$475 + \$30 = \$ 505 for the deviating agent.

However, this outcome is suboptimal from a social perspective. One socially optimal solution has one consumer adhere to their PST while the other deviates to slot 12, resulting in total costs of \$250 + \$0 and \$250 + \$300 respectively, for an aggregate cost of \$800.

Our research specifically emphasizes a decentralized approach, which calls for an examination of the Nash equilibrium solution where consumers are best responding to each other's actions.
Critically, the aggregated total cost of decentralized consumers (\$1000) remains higher than the socially optimal solution (\$800).
The socially optimal solution is not a Nash equilibrium, as one consumer can shift from the optimal action to secure a higher personal benefit.
This dynamic highlights the demand side load control problem as a social dilemma: the conflict arises from the trade-off between individual benefits and the social optimum, where individual actions aiming for personal gain result in suboptimal outcomes for society as a whole.

The RL baseline establishes that substantial inefficiency arises under decentralization in the full DSLM setting. However, the RL experiments do not provide a transparent explanation of equilibrium selection or entry dynamics: due to the nature of reinforcement learning agents, which operate in the presence of noisy gradients and require parameter tuning, validating the effects of payoff modifications on the utility landscape becomes challenging in the complex multi-appliance problem. We therefore turn to a minimal game abstraction to gain theoretical traction on the conditions under which cooperation can emerge.

\section{A minimal game abstraction}
\label{sec:minimal-game}

In its simplest form, the essential choice made by agents is to stick to their preferred appliance start times, or to move away from them in order to facilitate coordination.
We assume agents have two actions available: \PST\ stands for \emph{Preferred Start Time}; \Away\ indicates deviations from their most preferred times which results in a personal cost of inconvenience.

Considering a system with 2 consumers ($c_1, c_2$) each planning to schedule an appliance, we create a simple matrix to encapsulate the possible interactions (Table~\ref{tbl:2x2_normal_form_game}). Note that the entries in the matrix represent costs, so agents strive to minimize them.

\begin{table}[h!]
\centering
\begin{tabular}{|>{\columncolor{tableheadgray}}r |l|l|}
\hline
     & \cellcolor{tableheadgray}PST & \cellcolor{tableheadgray}Away \\ \hline
PST  & $1$, $1$ &    $0$, $1+p$      \\ \hline
Away & $1+p$, $0$ &  $2+p$, $2+p$    \\ \hline
\end{tabular}
\caption{Two-player, two-action normal-form game of load management. \label{tbl:2x2_normal_form_game}}
\end{table}

Scheduling on the PST without any additional demand from another agent is considered the optimal outcome, resulting in a cost of $0$. However, for the agent scheduling away, an additional inconvenience cost of $1+p$ is incurred. In the scenario where both agents schedule at the same time, a congestion cost of $1$ is imposed on each. For the costs in the matrix to represent a social dilemma, agents should prefer adherence to the \PST. In other words, moving \Away\ should not be the best response to \PST:
\begin{equation}
    1 < 1+p \implies 0 < p
\end{equation}
This condition makes (\PST, \PST) the only Nash equilibrium of the stage game. However, if (\PST, \PST) is an efficient equilibrium, this game would not be an interesting social dilemma; therefore, we additionally impose conditions ensuring that playing equilibrium (\PST, \PST) is not the social optimum:
\begin{equation}
    0+p+1 < 2 \implies p < 1
\end{equation}
Combining both inequalities yields the condition on $p$ that characterizes a load management game as a congestion game~\cite{milchtaich1996congestion}:
\begin{equation}
0 < p < 1
\label{eq:inequality1_for_q}
\end{equation}


We allow for this game to be repeated, which opens the door to coordination and cooperative turn taking.
Via backward induction, it is easy to show that if the game is repeated for a fixed number of rounds agents will only stick to their \PST\ as the only outcome supported in equilibrium.
Thus, following the literature on repeated games we assume an infinitely repeated game with a discount factor (continuation probability) of $0< \delta <1$.

\begin{figure}[ht]
    \centering
    \includegraphics[width=0.7\textwidth]{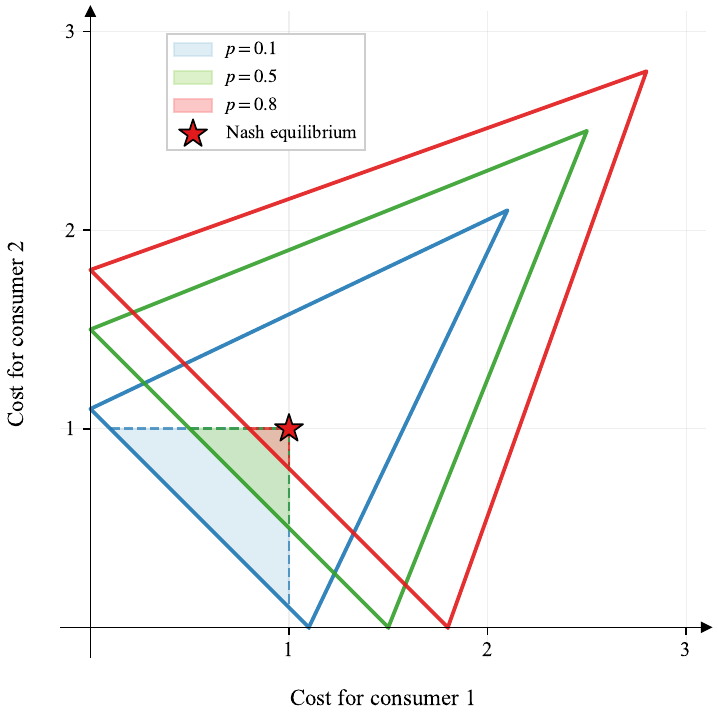}
    \caption{All possible equilibrium payoffs of the repeated games as the inconvenience cost $p$ varies. The shaded region illustrates mutually beneficial equilibrium payoffs; larger inconvenience costs shrink the space of cooperative equilibria.}
    \label{fig:folk_theorem}
\end{figure}

The number of strategies in these repeated games is uncountably infinite~\cite{garcia2016and}.
In the repeated prisoner's dilemma (with discounting) there is an infinite number of Nash equilibria.
This follows from the Folk theorem, which asserts that for large enough $\delta$, all payoff pairs in which both players get at least the mutual defection payoff can arise in equilibrium~\cite{67142936-4936-3f8a-be42-6f08b7a6c086}.
This means that we can expect cooperative as well as uncooperative outcomes. If we define cooperation as any deviation that benefits both parties involved, then the shaded area in Figure~\ref{fig:folk_theorem} corresponds to potential cooperative equilibria.

We restrict the strategy space by assuming agents only take into account the actions of their opponent in the most recent history period. The two actions of the opponent, $P$ representing \PST\ and $A$ representing \Away, give rise to $2$ potential states for the focal agent. By incorporating the initial state with no prior history, denoted as $S0$, the state space for each agent becomes $S = \{S0,P,A\}$. With an action space of $\{P,A\}$ a total of $2^{3}= 8$ deterministic memory-1 strategies can be identified.
The full payoff matrix for the eight deterministic memory-1 strategies is provided in Appendix~\ref{app:memory1-payoffs}.

Upon analysis, multiple pure Nash equilibria become evident. The scenarios (\textit{PPP},\textit{PPP}), (\textit{PPA},\textit{PPA}) and (\textit{PPP},\textit{PPA}) lead to agents initiating with \PST\ and consistently adhering to this choice. These Nash equilibria in the repeated game mirror those of the stage game and do not align with the socially optimal outcome. The remaining two equilibria emerge when agents alternate their actions, one initiating with \PST\ and the other with \Away, subsequently mirroring their opponent's actions for the remainder of the episode. These latter two equilibria (\textit{PPA}/\textit{APA}) result in substantially reduced costs for both players and can therefore be considered cooperative equilibria arising from successful coordination.
We restrict attention to these three strategies (PPP, PPA, APA) as they are the ones that emerge in Nash equilibria; the full set of payoffs justifying this restriction appears in Appendix~\ref{app:memory1-payoffs}.

For taking turns to be a beneficial alternative to \PST, the cost of \textit{APA} against \textit{PPA} should be lower than the cost of \textit{PPP} or \textit{PPA} against \textit{PPA}:
\begin{equation}
    \frac{(1+p)} {(1-\delta^2)} < \frac{1}{1-\delta} \implies p < \delta
\label{revised_ineq_q}
\end{equation}
Thus, successful coordination is only possible when the inconvenience cost is strictly less than the discount factor:
\begin{equation}
    0 < p < \delta
\label{revised_ineq_q_final}
\end{equation}

\section{Evolutionary dynamics and equilibrium selection}
\label{sec:evolutionary}

The Nash equilibrium represents a strategically stable outcome, where each player's strategy is optimal given the strategies chosen by the other players.
However, it alone does not reveal the strength of each equilibrium or indicate which ones are more likely to occur when agents are learning from rewards.
To address this aspect, we employ evolutionary dynamics.

We simplify the game, to allow for 3 possible strategies that capture the essential strategic elements.
These are the strategies that emerge in the Nash equilibria when considering all memory-1 strategies (\textit{PPP}, \textit{PPA}, \textit{APA}).

The replicator dynamics~\cite{nowak2006evolutionary} assumes a population of learning agents who tend to switch to better-performing strategies via social learning.
It assumes an infinitely large population~\cite{taylor1978evolutionary}, enabling us to represent the state of the system as a vector of numbers between $0$ and $1$, summing up to $1$:
\begin{equation}
    \dot{x_i} = x_i (\Pi_i - \bar{\Pi}),
\end{equation}
where $\dot{x_i}$ is the derivative of $x_i$, $\Pi_i$ is the average payoff of agents playing strategy $i$, and $\bar{\Pi}$ is the average payoff of the population.

The memory-1 strategy pairs involving turn-taking (\textit{PPA} and \textit{APA}) within the repeated load management game establish an asymmetric Nash equilibrium.
Thus the application of replicator dynamics as a single population is inadequate.
In the replicator dynamics for two populations~\cite{hofbauer1998evolutionary}, the assumption is that each player in one population engages in play with every player in the other population (but not within their population). For two populations $X$ and $Y$:
\begin{equation}
    \begin{split}
        \dot{x_i} = x_i((AY)_i-X^T \cdot AY) \\
        \dot{y_i} = y_i((AX)_i-Y^T \cdot AX)
    \end{split}
\label{eq_rep_2_pop}
\end{equation}

Given the symmetry of the game, we can represent it by a matrix:
\begin{equation}
    A = -1 \times
    \begin{bmatrix}
        1 & 1 & \delta  \\
        1 & 1 & \frac{\delta(1+p)}{1+\delta} \\
        (1+p)(1-\delta)+\delta & \frac{(1+p)}{1+\delta} & 2+p \\
    \end{bmatrix}
\label{nash_matrix}
\end{equation}

To investigate the size of the basins of attraction, we conduct Monte Carlo experiments randomizing the initial population that feeds into  replicator dynamics.
We maintain the parameter $p=0.5$, while adjusting $\delta$ to comprehend the shift in equilibria as $\delta$ approaches one.

\begin{figure}[ht]
    \centering
    \includegraphics[width=\textwidth]{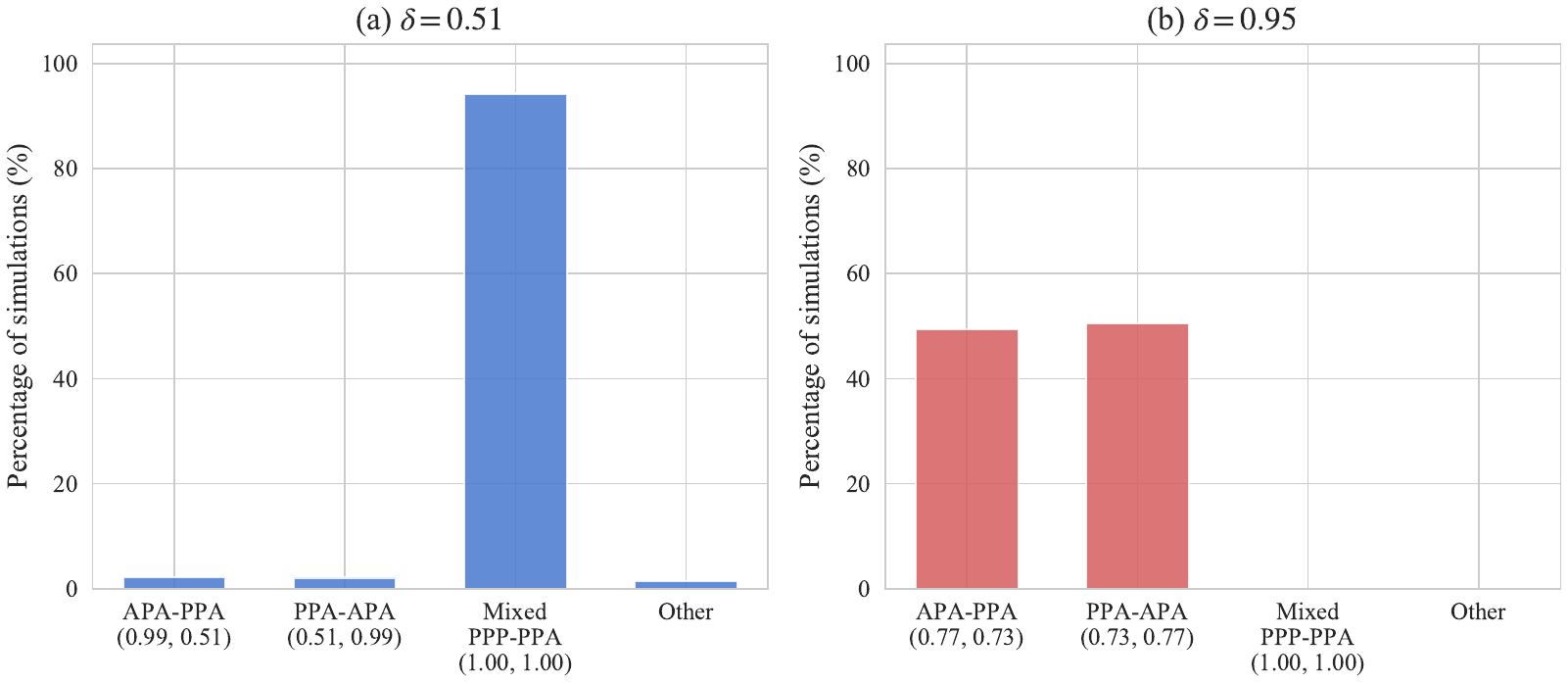}
    \caption{Convergence of $1000$ simulations of replicator dynamics with $p =0.5$ for two populations P1 and P2, comparing low versus high patience regimes. (a) For $\delta = 0.51$, the system predominantly converges to mixtures of PPP and PPA, resulting in non-cooperative outcomes. (b) For $\delta = 0.95$, a greater proportion of simulations converge to the cooperative turn-taking equilibria (PPA, APA).}
    \label{fig:replicator-hist}
\end{figure}

As observed in Figure~\ref{fig:replicator-hist}(a), for lower values of delta, the system tends to converge to a mixture of strategies \textit{PPP} and \textit{PPA}. Consequently, both agents consistently opt for \PST\ across all timesteps, resulting in non-cooperative behavior. Nevertheless, as $\delta$ approaches one, the stability of equilibria undergoes a shift, leading to a greater number of agent pairs converging towards cooperative equilibria (Figure~\ref{fig:replicator-hist}(b)). However, $\delta$ approaching one implies agents have infinite patience to collect their rewards, which is far from most real-world scenarios.

\section{Delegation to artificial agents via payoff shaping}
\label{sec:delegation}

When designing agents that will act on behalf of human principals, we can shape the reward function that guides their learning and decision-making.
We now examine an alternative method to enhance the stability of cooperative equilibria for lower $\delta$ values~\cite{GronauerDiepold2022CoopMARLSurvey,WillisEtAl2024MinimalRewardTransfer}.

\subsection{Intrinsic reward terms from globally observable signals}

We evaluate the impact of intrinsic reward terms on the utility landscape, analyzing how they contribute to the establishment of cooperation. To ensure that only cooperative behavior from both parties is rewarded, the intrinsic reward bonus is contingent upon the following three conditions:
\begin{itemize}
    \item Whether the focal agent has engaged in cooperation.
    \item Whether the focal agent's cost is lower than the average cost.
    \item Whether the population's cost is lower than the average population cost.
\end{itemize}

If the conditions are met, then the intrinsic reward term $I$ for the agent for the episode is calculated as:
\begin{equation}
    I = \frac{\Omega \times CoC}{R_E}
\end{equation}
where $\Omega$ is a large positive number, $CoC$ is the cost of cooperation, and $R_E$ is the episode reward of the agent.

The cost of cooperation is identified by comparing the focal agent's current schedule against the best response to the aggregate demand of others \cite{PereraDeNijsGarcia2025CoopEnsembles}. If the focal agent is not best responding and their deviation reduces prices for others, they have paid a cost of cooperation:
\begin{equation}
    \text{Cost of cooperation} = \text{Current cost} - \text{Cost of best response}
\end{equation}

Crucially, this payoff shaping scheme relies only on globally observable signals: aggregate demand and resulting prices.
The anonymous and diverse nature of opponents in the DSLM setting renders it considerably difficult to engage in any form of individual opponent modeling. Any attempts to model opponents are further complicated by the lack of knowledge regarding specific appliances, individual preferences, and inconveniences. With access to only the aggregate demand and the price, a simple way to postulate mutual cooperation is to identify situations that yield lower costs for both the focal agent and the rest of the population simultaneously.
In the full DSLM/RL environment, this is implemented as reward shaping; the concrete update rule is specified in Appendix~\ref{app:algo}.

\begin{figure}[ht]
    \centering
    \includegraphics[width=\textwidth]{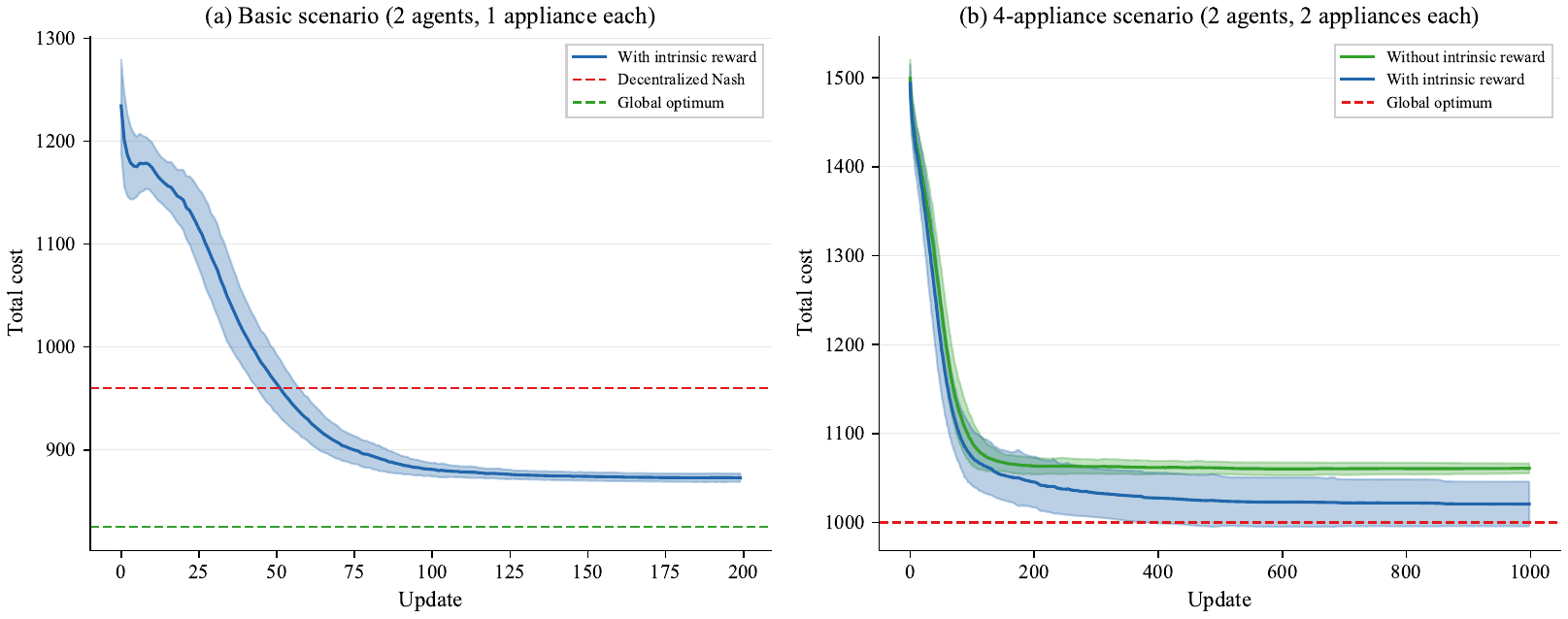}
    \caption{Average trajectory of system cost with intrinsic reward terms in the full DSLM/RL environment. (a) Basic scenario: agents adapt by alternating their actions, demonstrating turn-taking. (b) 4-appliance scenario: the intrinsic reward bonus enables agents to more frequently identify the socially optimal Nash equilibrium.}
    \label{fig:rl-intrinsic}
\end{figure}

As observed in Figure~\ref{fig:rl-intrinsic}(a), the total cost of agents experiences a notable drop as they adapt by alternating their actions, demonstrating a shared pattern of taking turns. In the 4-appliance scenario (Figure~\ref{fig:rl-intrinsic}(b)), the game presents two Nash equilibria, one of which aligns with the global optimum. When decentralized agents learn without the intrinsic reward term, they tend to gravitate towards the less desirable equilibrium from a social perspective. The introduction of the intrinsic reward bonus enables agents to more frequently identify the socially optimal Nash equilibrium. This empirical finding supports the theoretical claim that payoff shaping enlarges the basin of cooperative outcomes and improves equilibrium selection.

The new payoff matrix incorporating the intrinsic reward term is:
\begin{equation}
    A = -1 \times
    \begin{bmatrix}
        1 & 1 & \delta  \\
        1 & 1 & \frac{\delta(1+p)}{1+\delta} - \Omega(1-\delta)\\
        (1+p)(1-\delta) + \delta& \frac{(1+p)}{1+\delta}-\Omega(1-\delta) & 2+p \\
    \end{bmatrix}
\label{nash_matrix_2}
\end{equation}

The pure strategy equilibria remain consistent with those of the original game, but we observe a shift in the mixed-strategy (\textit{PPA}, \textit{APA}) equilibria toward more balanced ratios as $\Omega$ increases.

\begin{figure}[ht]
    \centering
    \includegraphics[width=\textwidth]{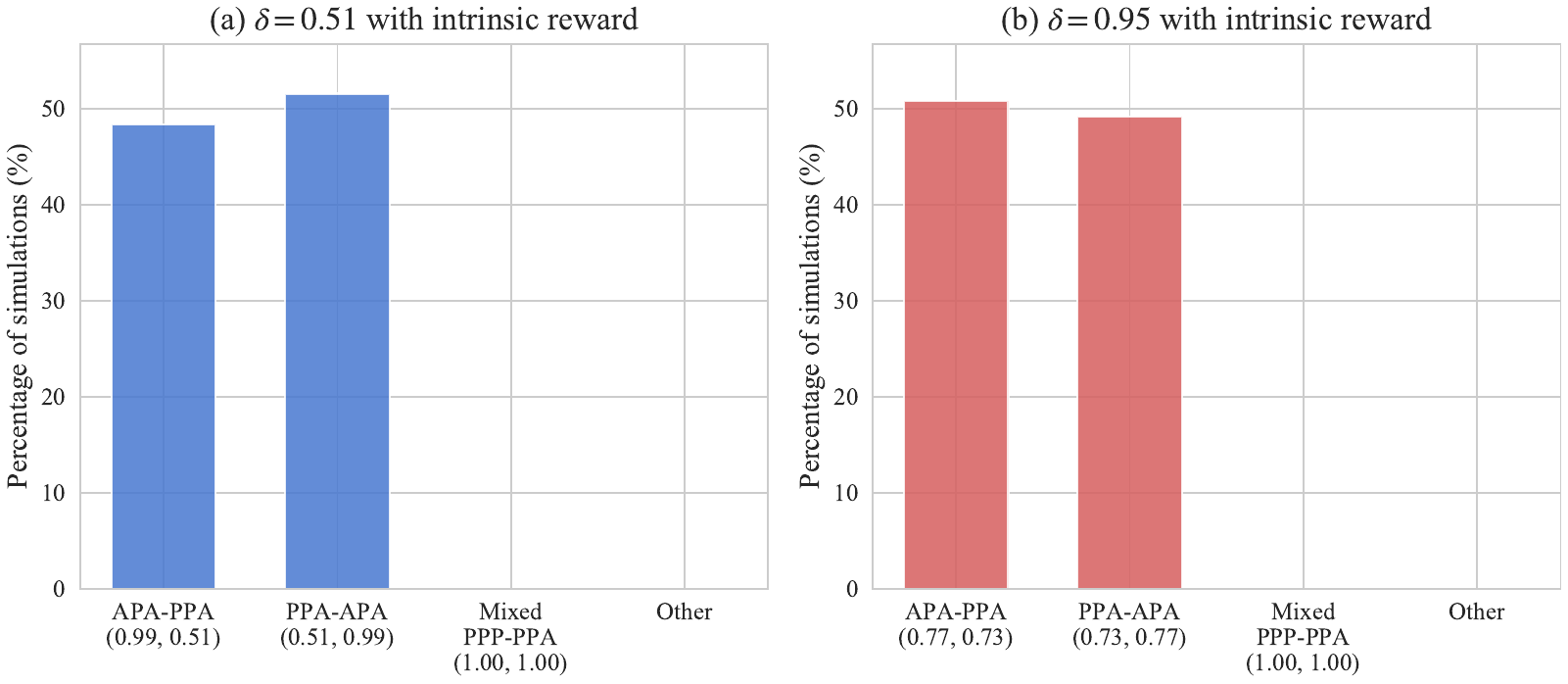}
    \caption{Convergence of $1000$ simulations of replicator dynamics with $p =0.5$ and intrinsic reward terms ($\Omega = 100$) for two populations P1 and P2. (a) For $\delta = 0.51$, the system converges to cooperative equilibria, in contrast to the non-cooperative outcome without intrinsic rewards (Figure~\ref{fig:replicator-hist}a). (b) For $\delta = 0.95$, cooperative convergence is similarly achieved.}
    \label{fig:with_intrinsic_delta_0.5}
\end{figure}

It is noticeable that even for lower $\delta$ values, the system converges to cooperative equilibria where one population converges to \textit{PPA} and the other to \textit{APA} (Figure~\ref{fig:with_intrinsic_delta_0.5}). Both populations are equally likely to end up in \textit{PPA} (or \textit{APA}) given the identical payoff matrices. This illustrates that, unlike the game without the intrinsic reward term, the game with payoff shaping can converge to desirable equilibria for lower delta values.

The intrinsic reward terms compel agents to converge towards cooperative strategies even for exceedingly low delta values (Figure~\ref{fig:cost_comparision}). In the specific scenario where $p=0.5$, agents can achieve a cost reduction of approximately $25\%$.

\section{Entry, partial adoption, and adoption resilience}
\label{sec:entry}

A central question for any cooperative scheme is whether early or partial adoption is viable.
In a free market setting, consumers will have to be convinced of the value of a load-managing device, with the option to choose not to participate in the scheme if they prefer. We use the intrinsic reward scheme to demonstrate that it possesses two advantageous properties: (1) it is a decentralized approach and (2) it is regret-free.

\begin{figure}[ht]
    \centering
    \begin{subfigure}[b]{0.48\textwidth}
        \centering
        \includegraphics[width=\textwidth]{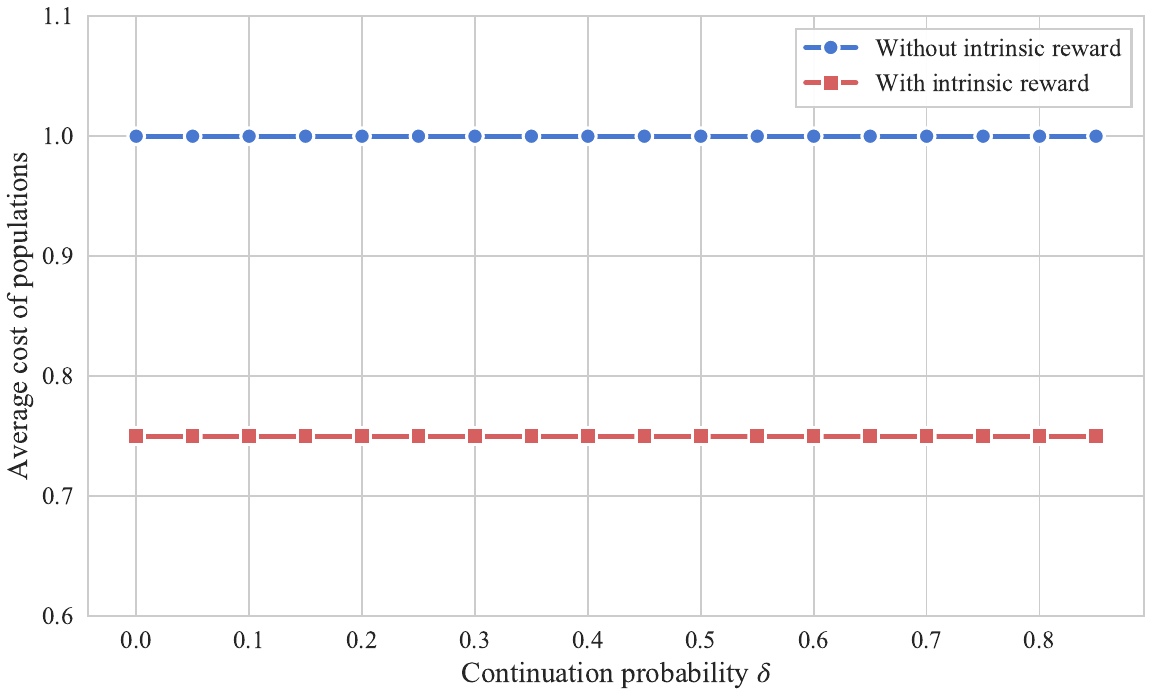}
        \caption{Symmetric adoption.}
        \label{fig:cost_comparision}
    \end{subfigure}
    \hfill
    \begin{subfigure}[b]{0.48\textwidth}
        \centering
        \includegraphics[width=\textwidth]{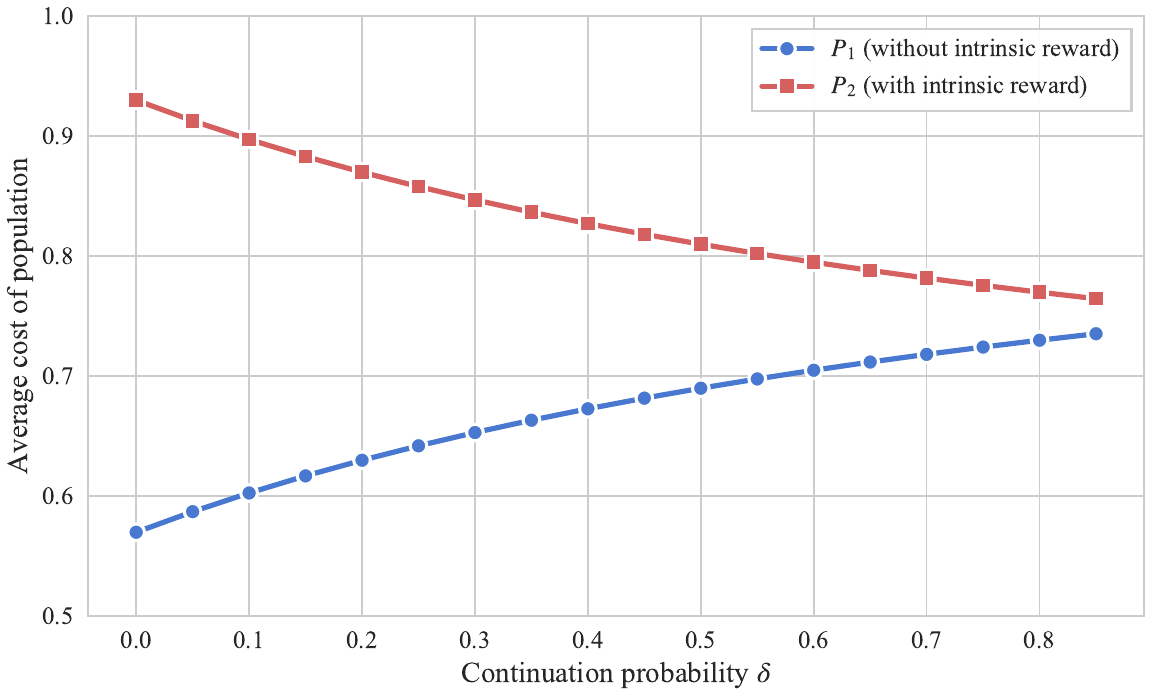}
        \caption{Asymmetric adoption.}
        \label{fig:resilience}
    \end{subfigure}
    \caption{Average population cost for varying $\delta$ ($\Omega = 100$, $p = 0.5$). (a) Both populations play the basic game (blue) or the game with intrinsic reward terms (red). (b) Population $P_1$ considers basic payoffs, $P_2$ considers payoffs with intrinsic reward terms.}
    \label{fig:cost_and_resilience}
\end{figure}

We aim to understand the behavior of agents using intrinsic reward terms when playing against agents adhering to the basic reward structure. We achieve this by using the payoff matrix for the basic game (Equation~\ref{nash_matrix}) for one population, \textit{P1}, and the updated payoff matrix (Equation~\ref{nash_matrix_2}) for the other, \textit{P2}. Figure~\ref{fig:resilience} illustrates the costs at convergence for each population at varying $\delta$ values.

As observed, for lower delta values, agents following the basic reward structure possess an advantage, since agents with modified rewards are more likely to initiate from \Away. However, when at least one agent follows the approach with intrinsic reward terms, it facilitates cooperation, subsequently reducing the cost for both individual agents compared to the scenario where both agents adhere to the basic game. Essentially, given the chance to adjust the payoffs, it is advantageous for a cost-minimizing agent to adopt this approach, irrespective of whether the opponent chooses to adapt or not. This property of resilience to non-adopters implies entry into the market is feasible.

However, when only one agent is adopting intrinsic reward terms, the non-adopter benefits more for lower continuation probabilities. This is a form of free-riding: non-adopters enjoy the benefits of cooperation induced by adopters without paying the cost of occasional deviation.

Despite this asymmetry, the key finding is that adopters are not structurally penalized.
Comparing Figure~\ref{fig:cost_comparision} with Figure~\ref{fig:resilience}, both adopters and non-adopters achieve lower costs than in the baseline non-cooperative outcome. Additionally, if a substantial proportion of market entries consist of such adopters, there is the potential for improved outcomes, even for non-adopters. In essence, the presence of adopters in the market, who are not disadvantaged by their adoption, creates an environment where the overall performance of the network can be enhanced.

Partial adoption therefore represents a viable path to improved social outcomes.
Users can choose to implement this payoff shaping scheme without facing extra costs due to non-adopters. The strategy is user-friendly, as it does not require the collective adoption of all consumers.

\section{Discussion and conclusion}
\label{sec:discussion}

We have studied the evolution of cooperation and coordination in a hybrid population where humans delegate strategic appliance scheduling to autonomous agents.
Our efforts were directed towards comprehending the demand-side load management problem by transforming it into a simplified version of a two-player scheduling game. We derived the conditions necessary for the DSLM problem to manifest as a mixed-motive game and conducted an evaluation of both the one-shot game and the repeated game.

Notably, the one-shot game yielded a single equilibrium that did not represent the globally optimal outcome. However, the dynamics shifted in the repeated game scenario, revealing the coexistence of globally favorable and unfavorable equilibria. The favorable equilibria involved agents taking turns scheduling at their preferred times without amplifying the aggregate demand, showcasing the potential for cooperation in repeated interactions.

In the pursuit of enhancing cooperative outcomes, we introduced intrinsic reward terms based on globally observable signals. Our analysis revealed that while the equilibria of the game remained consistent even with payoff shaping, the introduced intrinsic reward terms significantly influenced the strategies to which agents converged, particularly when continuation probability was lower. This observation is pivotal, considering that consumers, and consequently the agents representing them, do not possess infinite patience. The intrinsic reward terms guided agents towards cooperative equilibria, thereby reducing both collective and individual costs.

Furthermore, in a scenario involving non-adopting agents, the incorporation of intrinsic reward terms led to reduced costs for both parties, showcasing the resilience that this approach affords agents in the face of non-adopters. This implies that adopters can enter a market without facing a disadvantage.
However, our analysis also reveals that non-adopters may free-ride on the cooperation induced by adopters, particularly at lower discount factors. This asymmetry does not preclude beneficial entry, but it suggests that widespread adoption would further amplify social benefits.

Several limitations and future directions warrant acknowledgment.
The conditional intrinsic reward terms introduced will require further testing on more complex scenarios. As the agent's action space grows exponentially with the number of appliances within the system, conducting experiments becomes progressively resource-intensive. Additionally, challenges arise from the inability to synchronize agents' learning phases, potentially restricting the exploration of mutually cooperative actions. It is also noteworthy that these agents do not actively strive to encourage opponents to adopt cooperative equilibria, which presents an intriguing avenue for future investigation.

Future work includes: enhancing the identification of mutual cooperation with additional techniques, including statistical analysis of aggregate demand curves and daily prices; examining the performance of these agents in an environment with diverse other agents and heterogeneous preferences; and scaling the approach to larger populations while maintaining computational tractability.

In conclusion, the use of delegated agents with intrinsic reward terms presents a promising approach for addressing complex population games, particularly evident in domains like demand-side load management where individual information is limited, yet collective behavior can yield substantial benefits. By rewarding identified mutual cooperation conditionally, these agents can be directed toward cooperation while demonstrating resilience to non-adopters. Moreover, a significant portion of a population adopting such agents (not necessarily the entire population) is sufficient to lead agents towards more favorable equilibria.

\section*{Ethics}
This work did not require ethical approval.

\section*{Data Accessibility}
The simulation code and data supporting this article are available from the authors upon request.

\section*{Author Contributions}
I.P., F.d.N. and J.G. designed the research, performed the analysis, and wrote the manuscript.

\section*{Competing Interests}
The authors declare no competing interests.

\bibliographystyle{RS}
\bibliography{Bibliography}

\appendix

\section{Additional modelling and experimental details}
\label{app:details}

This appendix provides implementation details for the decentralized RL baseline in Section~\ref{sec:social-dilemma} and the intrinsic reward shaping used in Section~\ref{sec:delegation} of the main text.

\subsection{Full DSLM formulation}\label{sec:dslm}

We consider a set of consumers $i \in N$, each with their own set of appliances $j \in A_i$. An appliance is characterized by its activity duration $d_{i,j}$ and (constant) power consumption~$w_{i,j}$ while it is running.

All consumers' appliances must be scheduled to run (start \emph{and} finish) during a given planning period (e.g., one day), discretized into equal-length time steps (e.g., one hour), giving time~$t\in H$, $H = \{0, 1, \ldots, h\}$. The consumer provides three time points per appliance $0 \leq \textsc{est}_{i,j} \leq \textsc{pst}_{i,j} \leq \textsc{lst}_{i,j} \leq h - d_{i,j}$, the earliest, preferred, and latest start time respectively.

The goal of the DSLM problem is to minimize the total cost of electricity and the total appliance timing discomfort. We assume that the cost of electricity in a given time step is determined through \emph{merit order scheduling} of electricity generators. Electricity generators bid a (volume~$v_t$, price~$p_t$) pair, and the grid operator sorts these bids by increasing price, dispatching the generators up to the total demanded power~$P_t$, giving piecewise constant cost function
\begin{equation}
C_t(P_t) =
\begin{cases}
	p_{t,1} & \text{if } P_t \leq v_{t,1}, \\
	p_{t,2} & \text{if } v_{t,1} < P_t \leq v_{t,2}, \\
	\ldots \\
	p_{t,k} & \text{if } v_{t,k-1} < P_t \leq v_{t,k}.
\end{cases}
\end{equation}
To measure scheduling discomfort, consumers provide an appliance convenience sensitivity factor~$\eta_{i,j}$, measuring the (estimated financial) cost per time step that the appliance is scheduled away from~$\textsc{pst}_{i,j}$.

Given these cost coefficients, the DSLM problem can be formally specified as the following constrained optimisation problem:
\begin{equation}\label{eq:dslm}
\begin{aligned}
\min_{s_{i,j}}\: & \mathrlap{
					\sum_{i \in N}
					\Biggl(
						\sum_{t=0}^{h}\Bigl( C_t(P_t) \cdot P_{t,i}\Bigr) +
						\sum_{j \in A_i} \Bigl( \eta_{i,j} \cdot \left| s_{i,j} - \textsc{pst}_{i,j} \right| \Bigr)
					\Biggr)
				   } \\
\text{s.t. }     & \textsc{est}_{i,j} \leq s_{i,j} \leq \textsc{lst}_{i,j}		& \quad\forall i \in N, j \in A_i \\
			     & P_{t,i} = \sum_{j \in A_i} \Bigl( \mathbb{I}(s_{i,j} \leq t \wedge s_{i,j} + d_{i,j} > t) \cdot w_{i,j} \Bigr)  & \forall t \in H, i \in N \\
			     & P_t = \sum_{i \in N} P_{t,i}    & \forall t \in H
\end{aligned}
\end{equation}
The first constraint ensures that the appliance is scheduled at an allowed time. The second constraint uses an indicator function~$\mathbb{I}(.)$ to test if an appliance starting at $s_{i,j}$ is still running at time $t$, and if so, count its power consumption in the consumers' total power draw at time~$t$. The third constraint measures the total power consumption at time~$t$ by summing up the individual agents' consumption. The objective can be split up by agent into the total inconvenience cost~$I_{i}$ and per-time-step energy cost~$T_{i,t}$:
\begin{equation}
\begin{split}
	T_{i,t} &= C_t(P_t) \cdot P_{t,i}\\
	I_{i} &=\!\sum_{j \in A_i} \Bigl( \eta_{i,j} \cdot \left| s_{i,j} - \textsc{pst}_{i,j} \right| \Bigr) \\
\end{split}
\end{equation}
Crucially, each consumers' individual energy cost~$T_{i,t}$ depends on the cost factor~$C_t(P_t)$ which is influenced by all agents; in this way, the agents impose their externalities on others when they consume power during a given time.

The constraint problem as written down in Eq.~\eqref{eq:dslm} is a quadratic problem, however it can be linearized into a mixed-integer problem via the introduction of additional linking variables.

\subsection{Decentralized RL implementation}
\label{app:rl}

For the decentralized solution, we implement a decentralized RL algorithm treating each day as a single step. In each day, consumers (agents) decide the start time for each appliance based on the state provided and policy, $\pi$. The state is a vector with price table information of the day. We use Policy Gradient~\cite{sutton2018reinforcement} to update agent's policy $\pi$, by performing gradient ascent on the expected discounted reward with respect to the policy parameters.

\subsection{Intrinsic reward algorithm}
\label{app:algo}

\begin{algorithm}[H]
    \caption{Cost modification for decentralized learning}\label{alg:cost-mod}
    \begin{algorithmic}
    \State \textbf{Initialize:} $\Omega = \text{large constant}, A_{O} = A_{E} =\Omega, \text{update}= 0, m$
    \While{unconverged}
    \State $\text{update} \leftarrow \text{update} +1$
    \State \textbf{Initialize:} $ CoC \leftarrow 0 $
    \For{day in episode}
        \State $ R_B \leftarrow \text{cost of best response given others}$
        \If{$(R_E - R_B) > 0$}
            \State $ CoC \leftarrow CoC + R_E - R_B $
        \EndIf
    \EndFor
    \If{$(CoC>0) \land (R_O<A_O) \land (R_E<A_E) \land (A_E<P_A)$}
            \State $I_E \leftarrow \Omega \times (CoC/R_E)$
            \State $R_{\text{new}} \leftarrow \mu \times R_E - (1- \mu) \times I_E$
    \EndIf
    \State {$A_E \leftarrow \text{new episodic moving average cost}$}
    \State {$A_O \leftarrow \text{new episodic moving average price of others}$}
    \If{update \% m == 0}
    \State{$P_A \leftarrow A_E$}
    \EndIf
    \EndWhile
    \end{algorithmic}
\end{algorithm}

\subsection{RL training hyperparameters}
\label{app:hparams}

For the basic scenario with 2 agents and 1 appliance each, we used a batch size of 10,000 episodes, each spanning 2 days. Parameters were set as follows: $\Omega$ was set at 1500, $m$ was set to $1$, resetting the moving average $A_{E}$ with each update. The $\mu$ was assigned a value of 0.01, and the learning rate ($lr$) was set to 0.001.

For the scenario with 2 agents and 2 appliances each, we employed a batch size of 2,000 episodes, with each episode spanning 2 days. We set $\Omega$ to 500 and $m$ to $1$. $\mu$ was assigned a value of 0.01, and the learning rate ($lr$) was set to 0.001.

\subsection{Protocol behind the RL trajectory figure}
\label{app:rl-protocol}

The RL trajectory figure in the main text shows the average trajectory of system cost in the full DSLM/RL environment. The experimental protocol is as follows:

\begin{itemize}
    \item \textbf{Episode and step structure:} Each day is treated as a single step. An episode spans 2 days, during which agents select appliance start times based on observed price information.
    \item \textbf{Total cost:} The ``Total cost'' on the y-axis represents the sum of energy cost and inconvenience cost for the agent, as defined in Section~\ref{sec:social-dilemma}.
    \item \textbf{Baseline condition:} The baseline (without intrinsic reward terms) uses the standard Policy Gradient algorithm where agents optimize only their individual cost.
    \item \textbf{With intrinsic reward terms:} The intrinsic reward condition applies the cost modification scheme described in Algorithm~\ref{alg:cost-mod}, adding the intrinsic reward bonus $I$ when the three conditions (cooperation, below-average individual cost, below-average population cost) are satisfied.
    \item \textbf{Averaging:} Results are averaged over a sufficiently large number of runs.
\end{itemize}

\subsection{Data generation}

We follow He et al.~\cite{he2018fast} to synthesize problem instances based on real data to capture realistic consumer behavior. For each consumer appliance, we sampled the $PST$ from a distribution created by solar home electricity data in 2012--2013 by Ausgrid. The power requirements for the appliances are randomly selected from a list provided by Ausgrid, which includes commonly used appliances as of July 2015. The duration of each appliance is sampled using a Rayleigh distribution.

For each appliance, the earliest start time ($EST$) is selected randomly between 0 and the corresponding $PST$. To ensure that the latest start time ($LST$) falls within the planning slots, we randomly select $LST$ in a way that ensures $LST + \text{Appliance duration} < \text{Num planning slots}$. Additionally, an inconvenience factor is sampled randomly from a range of 0 to 10.

To generate bid-stacks, we utilize bid data obtained from the Australian Energy Market Operator (AEMO) spanning 396 days from July 2021 to July 2022. Roughly one-third of the days from each month are selected for testing purposes, while the remaining days are used to train the RL algorithm.

\subsection{Memory-1 strategy payoffs}
\label{app:memory1-payoffs}

The expected payoffs of the strategy \textit{PPA} playing against strategy \textit{APA}, denoted by $U(PPA,APA)$, can be simplified using geometric series:
\begin{equation}
    U(PPA,APA) = \lim_{n\to\infty} 0 + \delta \times (1+p) + \delta^2 \times 0 + \ldots + \delta^{(n-1)} \times 0+\delta^n \times (1+p)
\end{equation}
By multiplying both sides by $\delta^2$ and subtracting:
\begin{equation}
    U(PPA,APA) =  \frac{\delta \times (1+p) } {(1-\delta^2)}
\end{equation}
Similarly:
\begin{equation}
    U(APA,PPA) =  \frac{(1+p)} {(1-\delta^2)}
\end{equation}

\begin{sidewaystable}
    \tiny
    \centering
    \begin{tabular}{|l|l|l|l|l|l|l|l|l|}
    \hline
    \rowcolor{tableheadgray}
    $\Pi_1/\Pi_2$        & {\color{headerblue}$PPP$} & {\color{headerblue}$PPA$} & {\color{headerblue} $PAP$} & {\color{headerblue} $PAA$} & {\color{headerblue} $APP$} & {\color{headerblue} $APA$} & {\color{headerblue} $AAP$} & {\color{headerblue} $AAA$} \\ \hline

    \cellcolor{tableheadgray}{\color{rowlabelred}$PPP$} &
        \cellcolor{nashblue}\begin{tabular}[c]{@{}l@{}} $\frac{1}{1-\delta}$ \\ $\frac{1}{1-\delta}$ \end{tabular} &
        \begin{tabular}[c]{@{}l@{}} $\frac{1}{1-\delta}$ \\ $\frac{1}{1-\delta}$ \end{tabular} &
        \begin{tabular}[c]{@{}l@{}} $1$ \\ $1+ \frac{\delta(1+p)}{1-\delta}$ \end{tabular} &
        \begin{tabular}[c]{@{}l@{}} $1$ \\ $1+ \frac{\delta(1+p)}{1-\delta}$ \end{tabular} &
        \begin{tabular}[c]{@{}l@{}} $\frac{\delta}{1-\delta}$ \\ $1+p + \frac{\delta }{1-\delta}$ \end{tabular} &
        \begin{tabular}[c]{@{}l@{}} $\frac{\delta}{1-\delta}$ \\ $1+p + \frac{\delta }{1-\delta}$ \end{tabular} &
        \begin{tabular}[c]{@{}l@{}} $0$ \\ $\frac{1+p}{1-\delta}$ \end{tabular} &
        \begin{tabular}[c]{@{}l@{}} $0$ \\ $\frac{1+p}{1-\delta}$ \end{tabular} \\ \hline
    \cellcolor{tableheadgray}{\color{rowlabelred}$PPA$} &
        \begin{tabular}[c]{@{}l@{}} $\frac{1}{1-\delta}$ \\ $\frac{1}{1-\delta}$ \end{tabular} &
        \begin{tabular}[c]{@{}l@{}} $\frac{1}{1-\delta}$ \\ $\frac{1}{1-\delta}$ \end{tabular} &
        \begin{tabular}[c]{@{}l@{}} $\frac{1+\delta^2 (2+p)+\delta^3 (1+p)}{1-\delta^4}$ \\ $\frac{1+\delta (1+p)+\delta^2 (2+p)}{1-\delta^4} $ \end{tabular}&
        \begin{tabular}[c]{@{}l@{}} $1+ \frac{\delta^2 (2+p)}{1-\delta}$ \\ $1+\delta (1+p) + \frac{\delta^2(2+p)}{1-\delta}$ \end{tabular} &
        \begin{tabular}[c]{@{}l@{}} $\delta(1+p)+\frac{\delta^2}{1-\delta}$ \\ $(1+p)+\frac{\delta^2 }{1-\delta}$ \end{tabular} &
        \cellcolor{nashblue}\begin{tabular}[c]{@{}l@{}} $\frac{\delta(1+p)}{1-\delta^2}$ \\ $\frac{(1+p)}{1-\delta^2}$ \end{tabular} &
        \begin{tabular}[c]{@{}l@{}} $\frac{\delta(2+p)+\delta^2(1+p)+\delta^3}{1-\delta^4}$ \\ $\frac{(1+p)+\delta(2+p)+\delta^3}{1-\delta^4}$ \end{tabular} &
        \begin{tabular}[c]{@{}l@{}} $\frac{\delta(2+p)}{1-\delta}$ \\ $(1+p)+\frac{\delta(2+p)}{1-\delta}$ \end{tabular} \\ \hline
    \cellcolor{tableheadgray}{\color{rowlabelred}$PAP$} &
        \begin{tabular}[c]{@{}l@{}} $1+ \frac{\delta(1+p)}{1-\delta}$ \\ $1$ \end{tabular} &
        \begin{tabular}[c]{@{}l@{}} $\frac{1+\delta (1+p)+\delta^2 (2+p)}{1-\delta^4} $ \\ $\frac{1+\delta^2 (2+p)+\delta^3 (1+p)}{1-\delta^4}$ \end{tabular} &
        \begin{tabular}[c]{@{}l@{}} $\frac{1+\delta(2+p)}{1-\delta^2}$ \\ $\frac{1+\delta(2+p)}{1-\delta^2}$ \end{tabular} &
        \begin{tabular}[c]{@{}l@{}} $1+\delta(2+p)$ \\ $1+\delta(2+p)+\frac{\delta^2(1+p)}{1-\delta}$ \end{tabular} &
        \begin{tabular}[c]{@{}l@{}} $\delta +\frac{\delta^2(1+p)}{1-\delta}$ \\ $(1+p)+\delta  $ \end{tabular} &
        \begin{tabular}[c]{@{}l@{}} $\frac{\delta  +\delta^2(1+p)+\delta^3(2+p)}{1-\delta^4}$ \\ $\frac{(1+p)+\delta +\delta^3(2+p)}{1-\delta^4}$ \end{tabular} &
        \begin{tabular}[c]{@{}l@{}} $0$ \\ $\frac{1+p}{1-\delta}$ \end{tabular} &
        \begin{tabular}[c]{@{}l@{}} $0$ \\ $\frac{1+p}{1-\delta}$ \end{tabular} \\ \hline
    \cellcolor{tableheadgray}{\color{rowlabelred}$PAA$} &
        \begin{tabular}[c]{@{}l@{}} $1+ \frac{\delta(1+p)}{1-\delta}$ \\ $1$ \end{tabular} &
        \begin{tabular}[c]{@{}l@{}} $1+\delta (1+p) + \frac{\delta^2(2+p)}{1-\delta}$ \\ $1+ \frac{\delta^2 (2+p)}{1-\delta}$ \end{tabular} &
        \begin{tabular}[c]{@{}l@{}} $1+\delta(2+p)+\frac{\delta^2(1+p)}{1-\delta}$ \\ $1+\delta(2+p)$ \end{tabular} &
        \begin{tabular}[c]{@{}l@{}} $1+\frac{\delta(2+p)}{1-\delta}$ \\ $1+\frac{\delta(2+p)}{1-\delta}$ \end{tabular} &
        \begin{tabular}[c]{@{}l@{}} $\frac{\delta(1+p)}{1-\delta}$ \\ $(1+p)$ \end{tabular} &
        \begin{tabular}[c]{@{}l@{}} $ \delta(1+p) + \frac{\delta^2(2+p)}{1-\delta}$ \\ $(1+p) +\frac{\delta^2(2+p)}{1-\delta}$ \end{tabular} &
        \begin{tabular}[c]{@{}l@{}} $ \delta(2+p) + \frac{\delta^2(1+p)}{1-\delta}$ \\ $(1+p) +\delta (2+p) $ \end{tabular} &
        \begin{tabular}[c]{@{}l@{}} $\frac{\delta(2+p)}{1-\delta}$ \\ $(1+p)+\frac{\delta(2+p)}{1-\delta}$ \end{tabular} \\ \hline
    \cellcolor{tableheadgray}{\color{rowlabelred}$APP$} &
        \begin{tabular}[c]{@{}l@{}} $1+p + \frac{\delta }{1-\delta}$ \\ $\frac{\delta }{1-\delta}$  \end{tabular} &
        \begin{tabular}[c]{@{}l@{}} $(1+p)+\frac{\delta^2 }{1-\delta}$ \\ $\delta(1+p)+\frac{\delta^2}{1-\delta}$ \end{tabular} &
        \begin{tabular}[c]{@{}l@{}} $(1+p)+\delta  $  \\ $\delta  +\frac{\delta^2(1+p)}{1-\delta}$ \end{tabular} &
        \begin{tabular}[c]{@{}l@{}} $(1+p)$ \\ $\frac{\delta(1+p)}{1-\delta}$  \end{tabular} &
        \begin{tabular}[c]{@{}l@{}} $(2+p)+\frac{\delta }{1-\delta}$ \\ $(2+p)+\frac{\delta }{1-\delta}$ \end{tabular} &
        \begin{tabular}[c]{@{}l@{}} $(2+p)+ \frac{\delta^2}{1-\delta}$ \\ $(2+p)+\delta(1+p)+\frac{\delta^2}{1-\delta}$ \end{tabular} &
        \begin{tabular}[c]{@{}l@{}} $(2+p)+\delta  $ \\ $(2+p)+\delta +\frac{\delta^2(1+p)}{1-\delta}$ \end{tabular} &
        \begin{tabular}[c]{@{}l@{}} $(2+p)$ \\ $(2+p)+\frac{\delta(1+p)}{1-\delta}$ \end{tabular} \\ \hline
    \cellcolor{tableheadgray}{\color{rowlabelred}$APA$} &
        \begin{tabular}[c]{@{}l@{}} $1+p + \frac{\delta }{1-\delta}$ \\ $\frac{\delta }{1-\delta}$  \end{tabular} &
        \cellcolor{nashblue}\begin{tabular}[c]{@{}l@{}} $\frac{(1+p)}{1-\delta^2}$ \\ $\frac{\delta(1+p)}{1-\delta^2}$  \end{tabular} &
        \begin{tabular}[c]{@{}l@{}} $\frac{(1+p)+\delta+\delta^3(2+p)}{1-\delta^4}$ \\ $\frac{\delta  +\delta^2(1+p)+\delta^3(2+p)}{1-\delta^4}$ \end{tabular} &
        \begin{tabular}[c]{@{}l@{}} $(1+p) +\frac{\delta^2(2+p)}{1-\delta}$ \\ $ \delta(1+p) + \frac{\delta^2(2+p)}{1-\delta}$  \end{tabular} &
        \begin{tabular}[c]{@{}l@{}} $(2+p)+\delta(1+p)+\frac{\delta^2}{1-\delta}$ \\ $(2+p)+ \frac{\delta^2 }{1-\delta}$ \end{tabular} &
        \begin{tabular}[c]{@{}l@{}} $\frac{2+p}{1-\delta}$ \\ $\frac{2+p}{1-\delta}$ \end{tabular} &
        \begin{tabular}[c]{@{}l@{}} $\frac{(2+p)+\delta(1+p)+\delta^2}{1-\delta^4}$ \\ $\frac{(2+p)+\delta^2 +\delta^3(1+p)}{1-\delta^4}$ \end{tabular} &
        \begin{tabular}[c]{@{}l@{}} $\frac{2+p}{1-\delta}$ \\ $\frac{2+p}{1-\delta}$ \end{tabular} \\ \hline
    \cellcolor{tableheadgray}{\color{rowlabelred}$AAP$} &
        \begin{tabular}[c]{@{}l@{}} $\frac{1+p}{1-\delta}$ \\ $0$  \end{tabular} &
        \begin{tabular}[c]{@{}l@{}} $\frac{(1+p)+\delta(2+p) +\delta^3}{1-\delta^4}$  \\ $\frac{\delta(2+p)+\delta^2(1+p)+\delta^3}{1-\delta^4}$ \end{tabular} &
        \begin{tabular}[c]{@{}l@{}} $\frac{1+p}{1-\delta}$ \\ $0$  \end{tabular} &
        \begin{tabular}[c]{@{}l@{}} $(1+p) +\delta (2+p) $  \\ $ \delta(2+p) + \frac{\delta^2(1+p)}{1-\delta}$ \end{tabular} &
        \begin{tabular}[c]{@{}l@{}} $(2+p)+\delta +\frac{\delta^2(1+p)}{1-\delta}$ \\ $(2+p)+\delta  $  \end{tabular} &
        \begin{tabular}[c]{@{}l@{}} $\frac{(2+p)+\delta^2 +\delta^3(1+p)}{1-\delta^4}$ \\ $\frac{(2+p)+\delta(1+p)+\delta^2}{1-\delta^4}$  \end{tabular} &
        \begin{tabular}[c]{@{}l@{}} $\frac{(2+p)+\delta }{1-\delta^2}$ \\ $\frac{(2+p)+\delta }{1-\delta^2}$ \end{tabular} &
        \begin{tabular}[c]{@{}l@{}} $(2+p)$ \\ $(2+p)+\frac{\delta (1+p)}{1-\delta}$ \end{tabular} \\ \hline
    \cellcolor{tableheadgray}{\color{rowlabelred}$AAA$} &
        \begin{tabular}[c]{@{}l@{}} $\frac{1+p}{1-\delta}$ \\ $0$  \end{tabular} &
        \begin{tabular}[c]{@{}l@{}} $(1+p)+\frac{\delta(2+p)}{1-\delta}$ \\ $\frac{\delta(2+p)}{1-\delta}$  \end{tabular} &
        \begin{tabular}[c]{@{}l@{}} $\frac{1+p}{1-\delta}$ \\ $0$  \end{tabular}  &
        \begin{tabular}[c]{@{}l@{}} $(1+p)+\frac{\delta(2+p)}{1-\delta}$ \\ $\frac{\delta(2+p)}{1-\delta}$  \end{tabular} &
        \begin{tabular}[c]{@{}l@{}} $(2+p)+\frac{\delta(1+p)}{1-\delta}$ \\ $(2+p)$  \end{tabular} &
        \begin{tabular}[c]{@{}l@{}} $\frac{2+p}{1-\delta}$ \\ $\frac{2+p}{1-\delta}$  \end{tabular} &
        \begin{tabular}[c]{@{}l@{}} $(2+p)+\frac{\delta (1+p)}{1-\delta}$  \\ $(2+p)$ \end{tabular} &
        \begin{tabular}[c]{@{}l@{}} $\frac{2+p}{1-\delta}$ \\ $\frac{2+p}{1-\delta}$ \end{tabular} \\ \hline
    \end{tabular}
\caption{Payoffs of deterministic memory-1 strategies against each other. The Nash equilibria are highlighted in light blue.}
\label{tab:memory1-payoffs}
\end{sidewaystable}

\subsection{Jacobian stability analysis}

The ratios at a given equilibrium can be computed considering the stability of the point (equating derivatives $\dot{x_1},\dot{x_2},\dot{x_3},\dot{y_1},\dot{y_2}$ and $\dot{y_3}$ to zero), the constraints for proportions $x_1+x_2+x_3 =1$, $y_1+y_2+y_3 =1$, and the inequalities derived for parameters. The Jacobian is:
\begin{equation}
    J =
    \begin{pmatrix}
        \frac{\partial \dot{x_1}}{\partial x_1} & \frac{\partial \dot{x_2}}{\partial x_1} & \frac{\partial \dot{y_1}}{\partial x_1} & \frac{\partial \dot{y_2}}{\partial x_1}\\
        \frac{\partial \dot{x_1}}{\partial x_2} & \frac{\partial \dot{x_2}}{\partial x_2} & \frac{\partial \dot{y_1}}{\partial x_2} & \frac{\partial \dot{y_2}}{\partial x_2}\\
        \frac{\partial \dot{x_1}}{\partial y_1} & \frac{\partial \dot{x_2}}{\partial y_1} & \frac{\partial \dot{y_1}}{\partial y_1} & \frac{\partial \dot{y_2}}{\partial y_1}\\
        \frac{\partial \dot{x_1}}{\partial y_2} & \frac{\partial \dot{x_2}}{\partial y_2} & \frac{\partial \dot{y_1}}{\partial y_2} & \frac{\partial \dot{y_2}}{\partial y_2}
    \end{pmatrix}
    \label{J_appendix}
\end{equation}

The eigenvalues at the cooperative equilibria (\textit{APA}, \textit{PPA}) and (\textit{PPA}, \textit{APA}) are:
\begin{itemize}
    \item $e_1 = \frac{-2 - p - \delta}{1 + \delta}$
    \item $e_2 = e_3 = \frac{p - \delta}{1 + \delta}$
    \item $e_4 = \frac{(p - \delta) \delta}{1 + \delta}$
\end{itemize}
Given $0 < p < \delta$, all eigenvalues are negative, confirming stability of the cooperative equilibria.

\end{document}